\documentstyle[prb,aps,amsmath,psfig]{revtex}
\newcommand \be{\begin{eqnarray}}
\newcommand \ee{\end{eqnarray}}
\begin{document}
\twocolumn[\hsize\textwidth\columnwidth\hsize
           \csname @twocolumnfalse\endcsname
\title{Bernoulli potential in type-I and weak type-II
superconductors: III. Electrostatic potential above the vortex 
lattice}
\author{Pavel Lipavsk\'y$^{1,2}$, Klaus Morawetz$^{3,4}$, 
Jan Kol\'a\v cek$^2$, Ji{\v r}{\'\i} J. Mare{\v s}$^2$, 
Ernst Helmut Brandt$^5$ and Michael Schreiber$^3$}
\address{$^1$Faculty of Mathematics and Physics, Charles 
University, Ke Karlovu 5, 121 16, Prague 2, Czech Republic\\
$^2$Institute of Physics, Academy of Sciences, 
Cukrovarnick\'a 10, 16253 Prague 6, Czech Republic\\
$^3$Institute of Physics, Chemnitz University of Technology , 
09107 Chemnitz\\
$^4$Max-Planck-Institute for the Physics of Complex
Systems, Noethnitzer Str. 38, 01187 Dresden, Germany\\
$^5$Max-Planck-Institute for Metal Research,
         D-70506 Stuttgart, Germany}
\maketitle
\begin{abstract}
The electrostatic potential above the Abrikosov vortex lattice,
discussed earlier by Blatter {\em et al.} {[}PRL {\bf 77}, 566 
(1996){]}, is evaluated within the Ginzburg-Landau theory. 
Unlike previous studies we include the surface dipole. Close 
to the critical temperature, the surface dipole reduces the 
electrostatic potential to values below sensitivity of recent
sensors. At low temperatures the surface dipole is less 
effective and the electrostatic potential remains observable as 
predicted earlier.
\end{abstract}
    \vskip2pc]

\section{Introduction}
A boundary between the normal and superconducting states 
is characterized by two length scales. First, the local fraction of 
superconducting electrons is measured by the 
Ginzburg-Landau (GL) wave function $\psi$ which changes on 
the scale of the GL coherence length $\xi$. Second, there is a 
magnetic field $\bf B$ screened on the scale of the London penetration 
depth $\lambda$. This picture is common to type-I 
superconductors, in which the superconducting state is nearly separated from 
the normal state, and to type-II superconductors, in which the 
normal state is dispersed into individual lines called vortices.

In extreme type-II superconductors, where $\xi\ll\lambda$, the spatial shape 
of the superconducting fraction provides a sharper image of 
vortices than the spatial dependence of the magnetic field. Blatter 
{\em et al.\,}\cite{B96} proposed to observe the space 
modulation of the superconducting fraction via the electrostatic 
field that is expected to develop above a surface of 
superconductors. Their estimate of the electric field created by 
the Abrikosov vortex lattice predicts values well observable
by recent experimental tools.

The idea of Blatter {\em et al.\,} is as follows. According to 
theoretical predictions,\cite{H75,KK92,KF95} the space 
modulation of the superconducting gap $\Delta$ induces a 
charge transfer so that an electrostatic potential (called the 
Bernoulli potential) develops inside the superconductor. Since 
the Bernoulli potential $\phi$ is a function of the square of the 
gap, $\phi$ can be used to observe $|\Delta|^2$. The GL wave 
function is linearly proportional to the superconducting gap, i.e., 
$\phi$ can be used to observe $|\psi|^2$.

The Bernoulli potential cannot be detected inside the 
superconductor, it leaks out from the surface, however. Its
detection outside might be possible by scanning force 
microscopy or using the Kelvin capacitive pickup with a 
single-electron transistor as a sensor.\cite{B96}

As far as we know, such an experiment has not been performed 
yet, but it is under preparation. It is likely that the 
experimental setup will first be tested on conventional 
superconductors.\cite{set} In this paper we show that close to 
the critical temperature the electrostatic potential above the 
surface is strongly reduced by the surface dipole, which 
arises due to unbalanced pairing forces.\cite{p2} At lower 
temperatures the surface dipole is not so effective and the 
electrostatic field reaches observable amplitudes. 
 
The paper is organized as follows. In the next section we
specify the assumed experimental situation and the physical
picture of charge transfer contributing to the expected
electrostatic potential. In Sec.~III we derive a relation 
between the GL wave function and the electrostatic potential 
and provide a simple estimate of the potential for the magnetic 
field far from the upper critical field, i.e., for the limit of 
separated vortices. In Sec.~IV we discuss numerical results, 
and in Sec.~V we summarize.

\section{Basic assumptions}
Let us first describe the experimental situation we assume in 
our discussion. A superconducting film fills the layer $-L<z<0$. 
This film is thin on the characteristic scales of the GL theory, 
$L\ll\xi,\lambda$, but it is thick with respect to the BCS 
coherence length, $L\gg\xi_0$. The magnetic field ${\bf B}\| z$ 
penetrates the superconductor in form of vortices. The 
electrostatic potential will be scanned close to the surface at 
$z_{\rm scan}>0$.

Now we turn to the underlying physical picture. The 
electrostatic potential leaking out of the superconductor is 
generated by charges that can be sorted into three groups: 
{\bf A.} the bulk charge, {\bf B.} the surface dipole, and {\bf C.}
the surface charge. These contributions are discussed in 
individual subsections.

\subsection{Bulk charge}
The bulk charge covers a transfer of electrons from the inner 
to the outer regions of the vortices. There are various forces 
taking part in this transfer. First, electrons rotate around 
the vortex center so that the inertial force acts in the 
centrifugal direction. Second, the magnetic field pushes 
electrons via the Lorentz force -- also pointing outward. Third, 
the energy of Cooper pairs is lower than the energy of free 
electrons, therefore unpaired electrons in the vortex core are 
attracted towards the condensate around the core. The 
resulting force again points outward. These forces deplete the 
electron density in the vortex core, creating the Coulomb 
force which balances all the other forces.

Due to these various contributing mechanisms, there is no single
characteristic length scale of the charge modulation. The 
shortest scale is the GL coherence length $\xi$ reflecting the 
pairing forces. The contribution of the inertial and the Lorentz 
forces change on the scale of the London penetration depth 
$\lambda$. The long-range periodicity is of course enforced by 
the Abrikosov vortex lattice. 

The bulk charge has been evaluated within various 
approximations mostly covering only some of the acting forces. 
Studies based exclusively on the Lorentz and the inertial forces 
have been performed within the classical picture of the 
superconducting fluid \cite{LB97} and later with its quantum 
form.\cite{KLB01} 

In the last decade, the charge transfer in vortices has been 
derived from the electron-hole non-symmetry of the density of 
states at the Fermi level.\cite{KK92,KF95} Later it was 
recognized that the electron-hole non-symmetry effects are 
identical to forces due to the pairing energy.\cite{AW68,R69} 
In result there are two scientific dialects for the pairing forces. 
We prefer the original one.

Approaches that cover all the above listed forces are either 
phenomenological or microscopic. The phenomenological
studies\cite{LKMB02} use the theory of the GL type. The 
microscopic studies\cite{K01,MK02,MK03,MK03L,JG04} are based
on the Bogoliubov-de~Gennes theory. It is worth to mention 
mention that in the vicinity of $T_c$ the electrostatic potential 
obtained from the microscopic studies agrees with the result
of the GL theory.\cite{K01,MK02,MK03,MK03L,JG04} 

The microscopic theory is superior to the others as it covers
all important contributing mechanism in a unified way \cite{JG04} 
and its 
region of applicability is not restricted to the vicinity of $T_c$ 
or to small gradients of the superconducting gap. On the other
hand, such complete treatment is complicated and numerically
demanding. Here we use the simpler GL theory. 

\subsection{Surface dipole}
The second kind of charges determining the electrostatic
potential is a surface dipole due to which the potential has 
a finite step. A spatially resolved profile of this step is not known. It is expected 
that the dipole is located near the surface inside the 
superconductor. In analogy with the space profile of the 
superconducting gap, one can speculate that the width of the 
dipole is similar to the BCS coherence length $\xi_0$. The 
present treatment is limited to temperatures close to $T_c$, 
where $\xi_0\ll\xi$.  Accordingly, we assume the width of the 
surface dipole to be infinitesimal. 

So far, the role of the surface dipole in superconductors is not
fully clarified. There is an important experimental experience 
with the electrostatic potential above the surface in the 
Meissner state, however. Precise measurements of the 
electrostatic potential made by Morris and Brown\cite{MB71} 
with the help of the capacitive Kelvin method have shown a 
surprising result -- all contributions of the pairing 
forces\cite{AW68,R69} are canceled by the surface dipole. The 
experiment of Morris and Brown thus indicates that the surface dipole has to be taken into account. 

Based on the assumption that the surface dipole results 
from the surface depression of the BCS gap, we  have derived 
in Ref.~\onlinecite{p2} the local value of the dipole as a 
function of the GL wave function at the surface. The internal 
electrostatic potential and the potential step due to the 
surface dipole add. The resulting observable surface potential 
\begin{equation}
e\phi_0=-{f_{\rm el}\over n}
\label{s1}
\end{equation}
is given by the free energy per electron.\cite{p2} Here $n$ is 
the total density of pairable electrons (the total density) and 
$f_{\rm el}$ is the density of the electronic part of the free 
energy. 

The surface potential (\ref{s1}) follows from general 
thermodynamic assumptions and it can be implemented 
within different approximations of the free energy, compare 
Ref.~\onlinecite{p2} with Ref.~\onlinecite{LKMM01}. In our 
treatment the free energy is evaluated from the GL theory. 

\subsection{Surface charge}
Finally, there is the surface charge distributed on the scale of the Thomas-Fermi screening length, $\lambda_{\rm TF}$, 
from the surface. Since $\lambda_{\rm TF}$ is much shorter 
than the GL coherence length $\xi$, the BCS coherence length 
$\xi_0$, and the London penetration depth $\lambda$, we use 
the limit $\lambda_{\rm TF}\to 0$ and treat the surface charge 
as an ideal two-dimensional surface charge. The surface 
charge represents the utmost layer of the superconductor.

The surface charge simplifies the construction of the electrostatic 
potential inside and outside of the superconductor.\cite{B96} Neglecting 
contributions of the order of $\lambda_{\rm TF}^2/\xi^2$, one 
can simply evaluate the potential at the inner side of the surface, $\phi(0_-)$, and match it with the potential outside, 
$\phi(0_+)=\phi(0_-)$. The potential outside decays far from 
the surface to its mean value, $\phi(z)\to \langle\phi\rangle$ for $z\to\infty$. Due to this asymptotic condition, the potential is 
fully specified by its value at the surface. In our case the 
potential at the surface includes the surface dipole, 
$\phi(0_+)=\phi(0_-)=\phi_0$, where $\phi_0$ is given by 
formula (\ref{s1}). 

The matching of the inner and outer potentials is simple in the 
two-dimensional Fourier representation,
\begin{equation}
\phi({\bf K})={2\over \Omega}
\int_{\Omega}d{\bf r}\,\phi_0({\bf r})\cos({\bf Kr}),
\label{s3}
\end{equation}
where ${\bf r}\equiv (x,y)$ and ${\bf K}$ are discrete momenta 
that have to be selected according to the structure of the 
Abrikosov vortex lattice. The area $\Omega$ of the elementary 
cell is given by the mean magnetic field $B$ and the 
elementary flux, $\Phi_0=B\Omega$. One obtains the 
potential at any distance $z>0$ from the surface,
\begin{equation}
\phi({\bf r},z)=\langle\phi\rangle+\sum_{{\bf K}\ne 0}\phi({\bf K})
\,{\rm e}^{-|{\bf K}|z}\cos\left({\bf K}{\bf r}\right),
\label{s2}
\end{equation}
from the Fourier components (\ref{s3}) and the mean value 
$\langle\phi\rangle={1\over \Omega}\int d{\bf r}\,\phi_0({\bf r})$.
It is easy to check that the potential (\ref{s2}) satisfies the
Poisson equation.

The potential or its gradient at a finite distance will naturally be 
necessary for interpretations of future measurements. On the 
other hand, we feel that this technical step does not bring any 
new insight to the problem. In our discussion we will focus on 
the surface value $\phi_0({\bf r})=\phi({\bf r},0)$. 

\section{Surface potential within the Ginzburg-Landau theory}
With respect to electrostatic phenomena it is advantageous 
to introduce the GL theory in the formulation proposed by 
Bardeen.\cite{B54,B55} The free energy 
\begin{eqnarray}
f_{\rm el}&=&{1\over 2}\gamma T^2+
{1\over 2m^*}\bar\psi\left(-i\hbar\nabla-e^*{\bf A}
\right)^2\psi
\nonumber\\
&-&\varepsilon_{\rm con}{2|\psi|^2\over n}-
{1\over 2}\gamma T^2\sqrt{1-{2|\psi|^2\over n}}
\label{s4}
\end{eqnarray}
is composed of three terms: the free energy of Gorter and 
Casimir (last two terms), the kinetic energy in the quantum form 
(second term), and the subtracted free energy of the normal 
state (first term). Here, $\gamma$ is the linear coefficient of the 
specific heat. The condensation energy determines the critical 
temperature as $\varepsilon_{\rm con}={1\over 4}\gamma 
T_c^2$.

For temperatures close to $T_c$ the superconducting fraction 
is small, ${2|\psi|^2\over n}\ll 1$. Expanding the square root in 
this small value one arrives at the GL free energy
\begin{equation}
f_{\rm el}={1\over 2m^*}\bar\psi\left(-i\hbar\nabla-
e^*{\bf A}\right)^2\psi+\alpha|\psi|^2+{1\over 2}\beta
|\psi|^4.
\label{s4a}
\end{equation}
Note that the kinetic energy is not in the form proposed by 
Ginzburg and Landau\cite{GL50}. In particular, it becomes 
negative at the vortex core. The volume integral over this 
kinetic energy differs from the GL form by a surface 
contribution which is identically zero if the GL boundary 
condition is satisfied, see the discussion in 
Ref.~\onlinecite{W96}.

From the limit ${2|\psi|^2\over n}\to 0$ one obtains the GL 
parameters $\alpha=-{1\over 2n}\gamma\left(T_c^2-T^2\right)$ 
and $\beta={1\over 2n^2}\gamma T^2$. Below we compare the
surface potential with the bulk potential. The bulk potential
includes pairing forces proportional to density derivatives 
${\partial\alpha\over\partial n}$ and ${\partial\beta\over\partial 
n}$. The temperature dependence is essential in this point as
${\partial T_c\over\partial n}\ne 0$. 
After we have evaluated derivatives, we can use 
the usual limiting parameters of the GL theory, $\alpha=-
{1\over n}\gamma T_c\left(T_c-T\right)$ and $\beta={1\over 
2n^2}\gamma T_c^2$. 

To complete the description of the GL free energy we recall
that the charge of the Cooper pair equals twice the electron 
charge, $e^*=2e$. The mass $m^*=2 m$ depends on impurities 
and has to be fitted, e.g., from the GL coherence length using 
$\xi^2 m^*\gamma(T_c^2-T^2)=n\hbar^2$. For Niobium we use 
$T_c=9.5$~K and $\gamma=719$~J/(m$^3$K$^2$) giving $\varepsilon_{\rm con}=1.6\,10^4$~J/m$^3$. The electron 
density is $n=2.2\,10^{28}$/m$^3$, so that the condensation 
energy per particle is $\varepsilon_{\rm con}/n=
4.59$~$\mu$eV. For pure Niobium the effective mass is 
$m=1.2\,m_{\rm e}$ and the GL parameter is $\kappa=0.78$.

\subsection{The surface potential}
To proceed we have to derive the GL equation for the GL wave 
function $\psi$. For the assumed thin layer, induced currents 
are proportional to the layer width $L$ and they are negligible 
in the limit $L\ll\lambda$. The vector potential $\bf A$ thus has 
the same value as in the absence of the superconducting layer
and ${\bf B}$ is practically constant in thin layers \cite{B03}.

From the minimum of the free energy we arrive at the GL 
equation 
\begin{equation}
{\left(-i\hbar\nabla-e^*{\bf A}\right)^2\over 2m^*}\psi=-
\alpha\psi-\beta|\psi|^2\psi.
\label{s5}
\end{equation}
This equation is solved numerically with the help of the iteration 
procedure described in Ref.~\onlinecite{B97}. 

Before we enter a discussion of numerical results presented
in the next section, it is useful to express the surface potential 
as a function of the amplitude of the GL wave function. To this 
end we multiply (\ref{s5}) by $\bar\psi$ which yields
${1\over 2m^*}\bar\psi(-i\hbar\nabla-e^*{\bf A})^2\psi=-
\alpha|\psi|^2-\beta|\psi|^4$. If we substitute this relation into 
the free energy (\ref{s4a}), we find that the free energy is the
negative value of the quartic term $f_{\rm el}=-{1\over 2}
\beta|\psi|^4$. The surface potential (\ref{s1}) thus attains the
simple form
\begin{equation}
e\phi_0={1\over 2n}\beta|\psi|^4.
\label{s6a}
\end{equation}

At this point we can stress the first effect of the surface 
dipole. According to (\ref{s6a}), the electrostatic potential is 
proportional to $|\psi|^4$. In contrast, the internal potential
used by Blatter {\em et al.} 
\begin{equation}
e\phi_{\rm Bl}={\gamma T_c\over n}{\partial T_c\over\partial n}
|\psi|^2
\label{s6c}
\end{equation}
is proportional to $|\psi|^2$. The different order of the amplitude 
does not ruin the basic idea of the proposed experiment -- the 
electrostatic potential at the surface can be used to monitor the 
GL wave function.

\subsection{Upper estimate of the amplitude}
It is possible to establish an upper estimate of the amplitude 
of the electrostatic potential. The minimum of the potential 
(\ref{s6a}) is at a vortex center, where $e\phi_0\to 0$ since
$|\psi|^2 \to 0$. 

The potential $e\phi_0$ reaches its maximum somewhere 
between vortices. The magnetic field in the Abrikosov vortex 
lattice suppresses the amplitude of the GL wave function compared 
to its value $\psi_\infty^2=-{\alpha\over\beta}$ in the 
non-magnetic state. Therefore from $|\psi|^2\le \psi_\infty^2$ 
one obtains an upper estimate of the maximum as 
\begin{equation}
e\phi_0\le {1\over 2n}{\alpha^2\over\beta}=
{\varepsilon_{\rm con}\over n}\,4\,(1-t)^2.
\label{s6b}
\end{equation}
Here we have used 
$\alpha=(-4\varepsilon_{\rm con}/n)(1-t)$, where $t=T/T_c$, to 
highlight that close to the critical temperature the amplitude of 
the potential vanishes as $(1-t)^2$.

To illustrate how the surface dipole changes the surface 
potential we compare the full potential (\ref{s6a}) with the
internal potential (\ref{s6c}). For the amplitude of the internal 
potential (\ref{s6c}) one has the upper estimate
\begin{equation}
e\phi_{\rm Bl}\le -{\gamma T_c\over n}
{\partial T_c\over\partial n}
{\alpha\over\beta}={\varepsilon_{\rm con}\over n}
{\partial\ln T_c\over\partial\ln n}\,8\,(1-t) .
\label{s6d}
\end{equation}
For Niobium one finds from the McMillan formula the value
${1\over 2}\gamma T_c{\partial T_c\over\partial n}=
6.78$~$\mu$eV, see Ref.~\onlinecite{LKMB02}. This 
corresponds to ${\partial\ln T_c\over\partial \ln n}=0.74$. 
Numerical factors of the surface potential (\ref{s6b}) and the 
internal potential (\ref{s6d}) are thus quite comparable. What 
makes the difference is the temperature dependence. 

\subsection{Lower temperatures}
The above upper estimate of the potential amplitude is based 
on the zero free energy at the vortex center and the finite free 
energy in the non-magnetic state. These values can be easily
estimated also at temperatures far from $T_c$. Indeed, as long
as the free energy is adjusted to vanish in the normal state, it 
has to be zero at the vortex center, where the superconducting 
condensate vanishes. The free energy in the non-magnetic 
state can be taken from the thermodynamic measurements in 
the form $f_{\rm el}=-\varepsilon_{\rm con}(1-t^2)^2$. 
Accordingly, the upper estimate of the amplitude of the 
potential is
\begin{equation}
e\phi_0\le {\varepsilon_{\rm con}\over n} \,(1-t^2)^2.
\label{s6e}
\end{equation}
This estimate applies to any temperature. Of course, it 
approaches (\ref{s6b}) for $t\to 1$.

Similarly we can use the relation between the GL wave function and
the superconducting density $2|\psi|^2=n_s$ together with the
phenomenological law $n_s=n\,(1-t^4)$ to obtain the estimate of
the internal potential
\begin{equation}
e\phi_{\rm Bl}\le {\varepsilon_{\rm con}\over n}
{\partial\ln T_c\over\partial\ln n}\,2\,(1-t^4) .
\label{s6f}
\end{equation}
For $t\to 1$ this estimate goes to (\ref{s6d}).

One can see that for low temperatures, $t\ll 1$, estimates 
(\ref{s6e}) and (\ref{s6f}) yield comparable amplitudes of the 
potential. This indicates that at low temperatures the surface 
dipole is less effective in reducing the surface potential.

\section{Numerical results for Niobium}
Now we discuss the electrostatic potential with the help of 
numerical results. To be specific, we use material parameters of 
Niobium.

\begin{figure}
\psfig{file=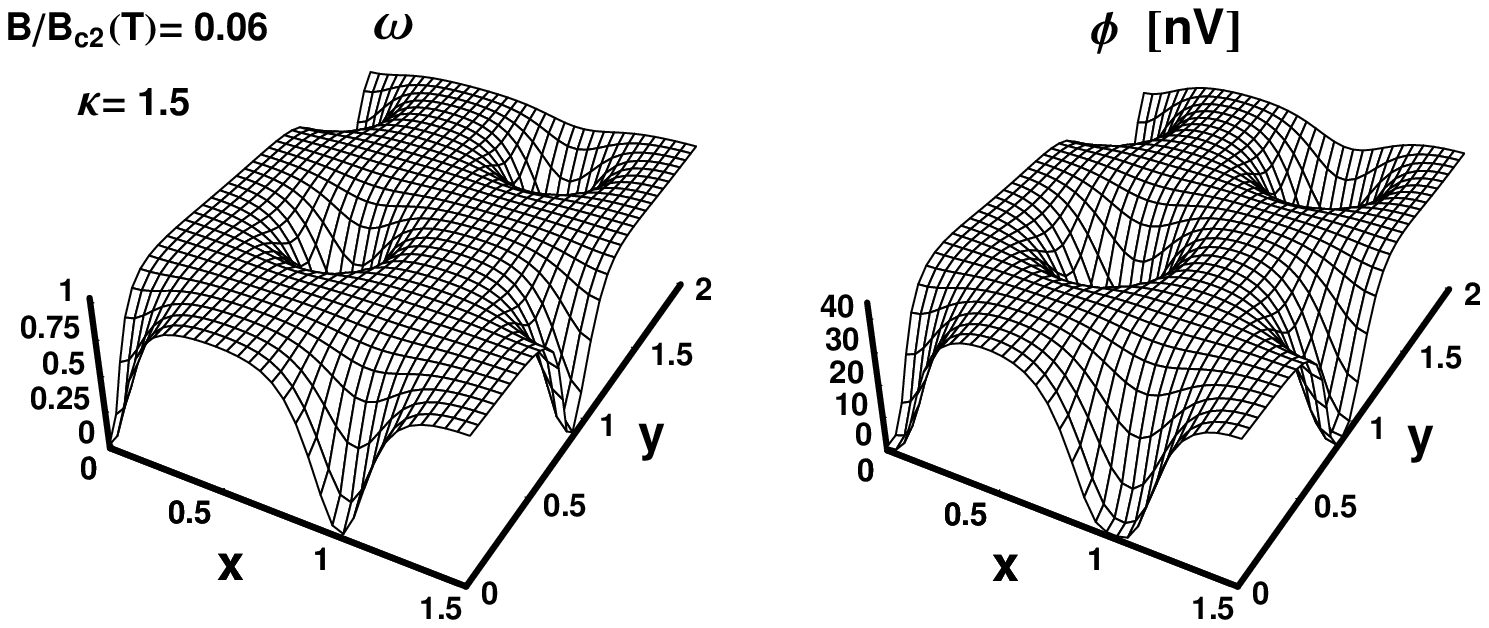,width=18cm}
\vskip -3mm
\psfig{file=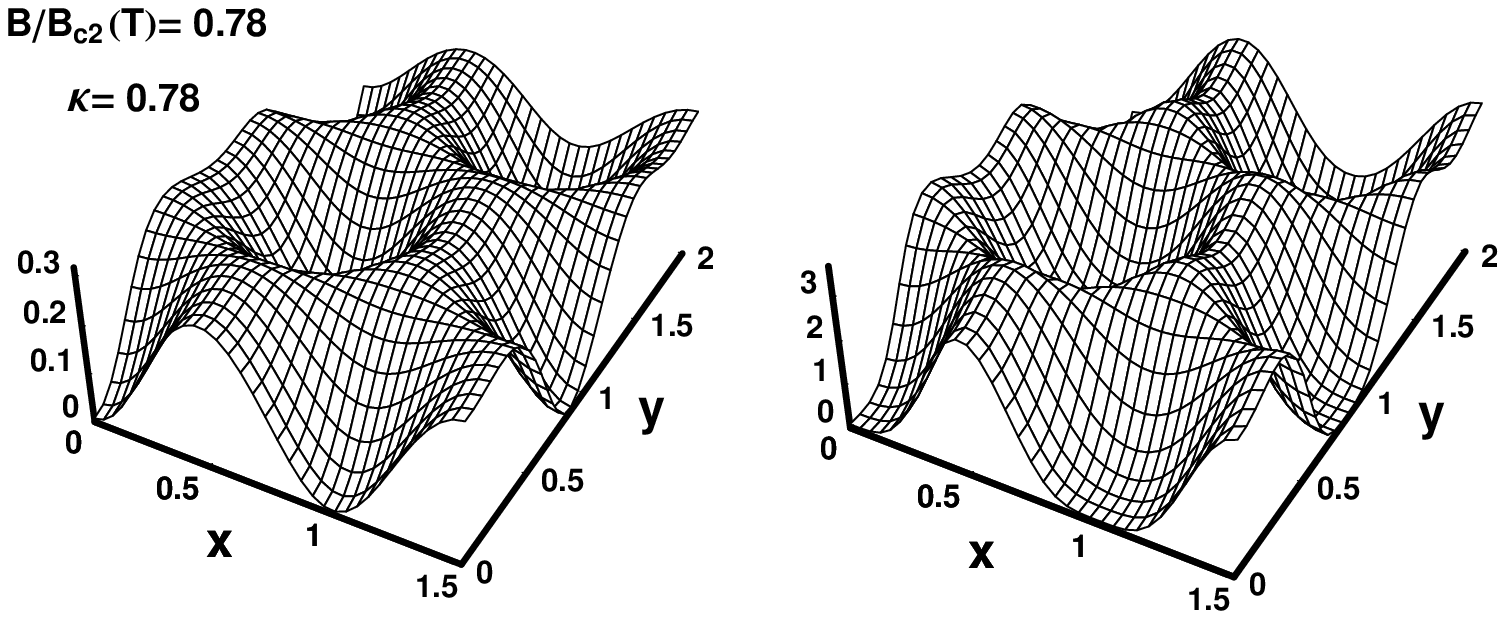,width=18cm}
\caption{The relative superconducting fraction $\omega=|\psi|^2/
\psi_\infty^2$ (left column) and the electrostatic potential $\phi$ 
(right column) at the surface of the superconductor with the 
Abrikosov vortex lattice. The distances are normalized to the vortex separation. The temperature $T=0.95\,T_c$ is used for
both cases, the GL parameter $\kappa$ and the magnetic field 
are specified for each row. Material parameters are of Niobium.
\label{fig1}}
\end{figure}

Figure~\ref{fig1} displays the GL wave function 
$\omega\equiv |\psi|^2/\psi_\infty^2$, where $\psi_\infty$ is 
the GL wave function in the absence of the magnetic field, 
$2\psi_\infty^2=n(1-t^4)$. The profile of $\omega$ is 
compared with the surface electrostatic potential 
$\phi_0$ according to equation (\ref{s6a}) near the critical 
temperature, $T=0.95\,T_c$.

In the first row we show the result for Niobium with
$\kappa=1.5$. The low magnetic field $B=0.06\,B_{c2}$ 
already falls into the limit of isolated 
vortices, because vortex cores are well separated and the 
superconducting fraction between them reaches its 
non-magnetic value with less than 1\% difference. The amplitude
of the electrostatic field is thus well described by the estimate
(\ref{s6e}).

Compared to the superconducting fraction, the electrostatic 
potential is much flatter in the center of the vortex. This feature 
directly follows from formula (\ref{s6a}). At the vortex center, 
$x^2+y^2=r^2\to 0$, the superconducting fraction quadratically 
vanishes with r, $\omega\propto r^2$. According to (\ref{s6a}) the 
potential vanishes there with the fourth order of $r$, $\phi_0\propto 
r^4$. We believe that this will be possible to be observed in future
measurements.

The second row in figure~\ref{fig1} with $\kappa=0.78$ corresponds to pure
Niobium. We choose the magnetic field close to the upper 
critical field $B=0.7818~B_{c2}$, when the superconducting 
fraction is suppressed to less than 1/3 of its non-magnetic 
value. In this regime the GL wave function is well approximated 
by the asymptotic solution due to Abrikosov.\cite{G66}

\onecolumn
\twocolumn 

\subsection{Estimate from the Abrikosov solution}
For $B$ close to $B_{c2}$, the amplitude of the GL wave 
function undergoes rapid changes while its shape remains
nearly unchanged. As in Fig.~\ref{fig1} we express the
superconducting fraction, $2|\psi|^2/n$, in terms of its 
non-magnetic value, $2|\psi_\infty|^2/n=1-t^4$ as
the relative superconducting fraction $\omega=|\psi|^2/|\psi_\infty|^2$.

For the fixed shape of the GL wave function, the mean 
$\langle\omega\rangle={1\over\Omega}\int 
d{\bf r}~\omega$ and its fluctuation $\langle\omega^2\rangle=
{1\over\Omega}\int d{\bf r}~\omega^2$ are proportional as 
$\langle\omega^2\rangle=\beta_{\rm A}\langle\omega
\rangle^2$, where the Abrikosov coefficient for the triangular 
lattice is $\beta_{\rm A}=1.16$. Taking $1-b$ with $b=B/B_{c2}$ 
as a small perturbation, one finds that the free energy 
(\ref{s4a}) has the minimum when $\langle\omega^2\rangle=
\langle\omega\rangle\left(1-b\right)$. The mean and the fluctuations thus read\footnote{For details 
of the derivation see Ref.~\onlinecite{G66}. In the limit of the 
thin layer the term $\langle\omega^2\rangle/2\kappa^2$ 
disappears as it follows from the induced magnetic field.} 
\begin{equation}
\langle\omega\rangle={1\over\beta_{\rm A}}
\left(1-b\right),~~~~~~~~
\langle\omega^2\rangle={1\over\beta_{\rm A}}
\left(1-b\right)^2.
\label{s6g}
\end{equation}

With the help of the mean values (\ref{s6g}) we can express the
mean values of the electrostatic potentials. The surface potential (\ref{s6a}) with the surface dipole included has the
mean value
\begin{equation}
\langle e\phi_0\rangle={\varepsilon_{\rm con}\over n}
\left(1-t^2\right)^2{1\over\beta_{\rm A}}
\left(1-b\right)^2.
\label{s6h}
\end{equation}
If we extend the estimate by Blatter {\em et al.} to magnetic 
fields close to the critical field, we find that it has the mean 
value
\begin{equation}
\langle e\phi_{\rm Bl}\rangle={\varepsilon_{\rm con}\over n}
{\partial\ln T_c\over\partial\ln n}\,2\,(1-t^4)
{1\over\beta_{\rm A}}\left(1-b\right) .
\label{s6i}
\end{equation}

Comparing (\ref{s6h}) with (\ref{s6i}) one can see that due to 
the surface dipole, the electrostatic potential vanishes close to 
the upper critical field as $(1-b)^2$, rather than $1-b$ without 
the dipole. We expect that this dependence on the magnetic 
field might be one of experimentally well accessible tests of the 
presence of the surface dipole.

\subsection{Effect of the surface dipole}
As already mentioned the surface dipole is responsible for the 
flat region of the potential at the center of the vortex. In 
this section we discuss the role of the surface dipole in more 
detail.

\begin{figure}
\psfig{file=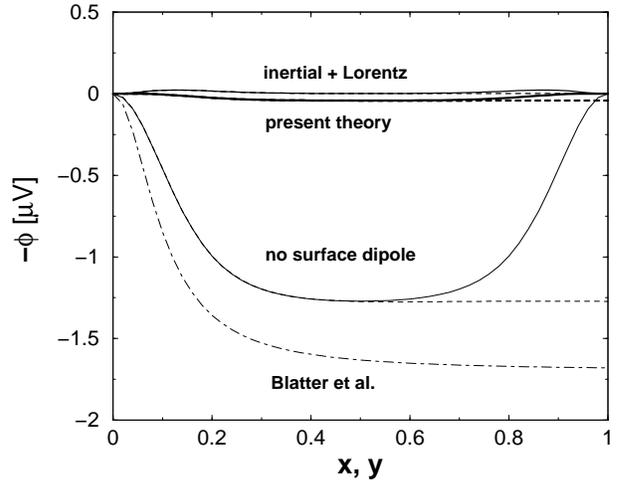,width=8cm,angle=-90}
\caption{Various approximations of the electrostatic potential 
at the surface. The parameters $T=0.95~T_c$, $\kappa=1.5$ 
and $B=0.7818~B_{c2}$ correspond to the upper right plot in 
Fig.~\protect\ref{fig1}. The symmetric lines (solid) are cuts 
along the $x$ axis, the non-symmetric lines (dashed) are cuts
along the $y$ axis in Fig.~\protect\ref{fig1}. The line `present 
theory' is given by eq. (\protect\ref{s6a}), the line `no surface 
dipole' by eq. (\protect\ref{chi6}), the line `inertial $+$ Lorentz' 
corresponds to the first term of eq. (\protect\ref{chi6}), and the 
line `Blatter et al.' is according to Ref.~\protect\onlinecite{B96}.
\label{fig2}}
\end{figure}

In Fig.~\ref{fig2} we compare the present theory yielding 
formula (\ref{s6a}) with three approximations. First, if one
neglects the surface dipole, the surface potential becomes
equal to the internal potential\cite{LKMB02} 
\begin{eqnarray} 
e\phi &=&-{1\over 2m^*n}
\bar\psi\left(-i\hbar\nabla-e^*{\bf A}\right)^2\psi
\nonumber\\
&+&{\partial\varepsilon_{\rm con}\over\partial n}
{2|\psi|^2\over n}-{T^2\over 2}{\partial\gamma\over\partial n}
\left({|\psi|^2\over n}+{|\psi|^4\over 2n^2}\right).
\label{chi6}
\end{eqnarray}
The material parameters for Niobium 
${\partial\varepsilon_{\rm con}\over\partial n}=8.71$~$\mu$eV 
and ${1\over 2}{\partial\gamma\over\partial n}T_c^2=
3.85$~$\mu$eV, are derived in Ref.~\onlinecite{LKMB02}.

Second, the surface dipole results from the pairing forces. 
As long as one does not account for the surface dipole,
perhaps it is better to neglect also other pairing forces. In 
this approximation the surface potential equals the first
term of equation (\ref{chi6}) which covers the inertial and 
Lorentz forces.

Third, following Blatter {\em et al.} we take a single vortex 
located at $x,y=0$. Its GL wave function is modeled by the Clem 
ansatz\cite{C75} $\omega\approx 1-\xi_{\rm v}^2/(x^2+y^2+
\xi_{\rm v}^2)$. The vortex radius $\xi_{\rm v}$ found from the 
minimum of the free energy is given by $\xi_{\rm v}=\xi\sqrt{2}
\sqrt{1-K_0^2(\xi_{\rm v}/\lambda)/K_1^2(\xi_{\rm v}/\lambda)}$, 
where $K_0$ and $K_1$ are modified Bessel functions. 
According to the approximation of Khomskii and 
Kusmartsev\cite{KK92} adopted by Blatter {\em et al.}, we take 
only the second term of equation (\ref{chi6}) with ${\partial
\varepsilon_{\rm con}\over\partial n}\approx{1\over 2}\gamma 
T_c{\partial T_c\over\partial n}=6.78$~$\mu$eV.

The potentials plotted in Fig.~\ref{fig2} can be sorted into two 
groups. The internal potential 
(\ref{chi6}) and Blatter's result are very similar except 
for some minor differences following from the Clem model and 
the neglect of ${\partial\gamma\over\partial n}$. The 
potential (\ref{s6a}) with the surface
dipole included, and the approximation by inertial and Lorentz 
forces are much smaller than the 
potentials from the first group.

\begin{figure}
\psfig{file=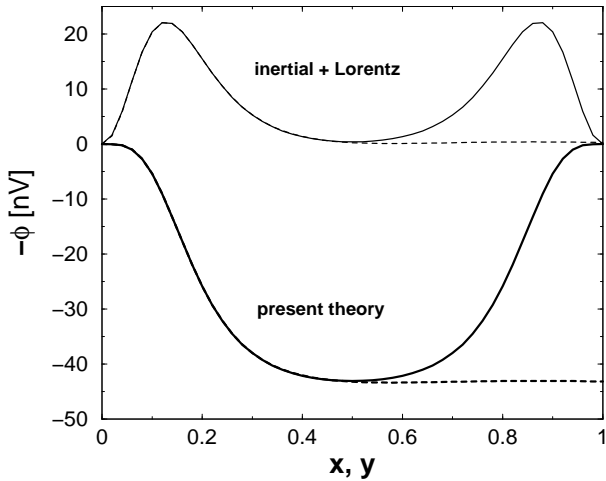,width=8cm,angle=-90}
\caption{Detail of Fig.~\ref{fig2}.\label{fig3}}
\end{figure}

To compare our potential (\ref{s6a}) with the approximation by 
inertial and Lorentz forces, we have expanded the scale in 
Fig.~\ref{fig3}. The numerical result clearly shows that the 
approximation by inertial and Lorentz forces has a very
different profile. Briefly, above the Abrikosov vortex lattice the
surface dipole cancels the major part of the contribution of
pairing forces. On the other hand, it is not possible to avoid the 
full calculation and to replace the action of the surface dipole 
by simply omitting the pairing forces.

\begin{figure}
\psfig{file=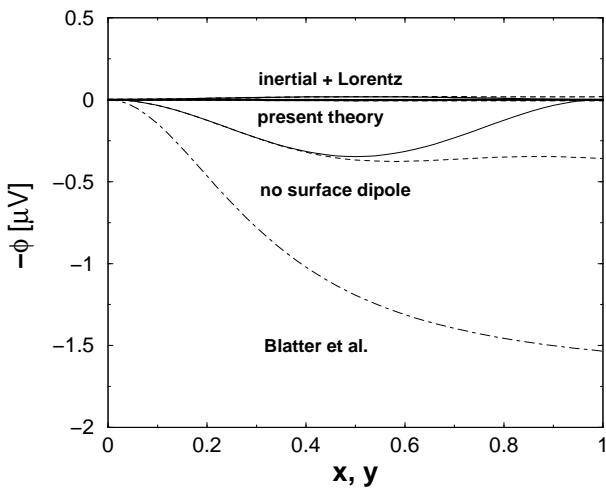,width=8cm,angle=-90}
\caption{The same as Fig.~\ref{fig2} but with parameters of
the lower row in Fig.~\protect\ref{fig1}, i.e., $T=0.95\,T_c$,  
$\kappa=0.78$ and $B=0.78~B_{c2}$.\label{fig4}}
\end{figure}

Figure~\ref{fig4} shows the different potentials for $\kappa=0.78$ corresponding to pure Niobium of the lower row in Fig.~\ref{fig1}. Most of the features are 
identical to the situation of larger $\kappa$ in the upper row in 
Fig~\ref{fig1} and in Fig.~\ref{fig2}. 
But, in the internal potential (\ref{chi6}) the GL wave 
function is now suppressed by the 
magnetic field compared to
Blatter's approach.
Indeed, the Clem approximation is derived for
the limit of low magnetic fields and thus it does not cover the 
suppression.

\begin{figure}
\psfig{file=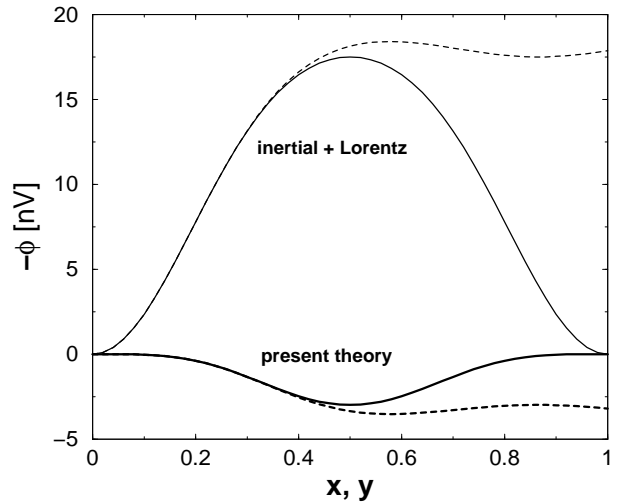,width=8cm,angle=-90}
\caption{Detail of Fig.~\ref{fig4}.\label{fig5}}
\end{figure}

The detail presented in Fig.~\ref{fig5} demonstrates that the full 
theory and the approximation by inertial and Lorentz forces 
result in very different profiles of the electrostatic potential. One 
can see that the neglect of pairing forces leads to the potential 
of much larger amplitude and the wrong sign. 

\section{Summary}
We have evaluated the electrostatic potential above the 
surface of a thin superconducting layer with the Abrikosov 
vortex lattice. It has been shown that the surface dipole 
strongly modifies the magnitude of this potential, in particular
when the GL wave function has a small magnitude. This is
due to the relation $\phi_0\propto |\psi|^4$, see (\ref{s6a}),
while without the dipole one finds $\phi_{\rm Bl}\propto |\psi|^2$.

According to various mechanisms to suppress the GL wave
function, we can outline possible cases for which the presented
theory can be tested. At the vortex core $|\psi|^2\propto r^2$
so that $\phi_0\propto r^4$ while $\phi_{\rm Bl}\propto r^2$.
At temperatures close the critical temperature, $t\to 1$, 
$|\psi|^2\propto 1-t$, therefore $\phi_0\propto (1-t)^2$ while
$\phi_{\rm Bl}\propto 1-t$. Finally, at magnetic fields close to
the upper critical field, $b\to 1$, $|\psi|^2\propto 1-b$ so that
$\phi_0\propto (1-b)^2$ while $\phi_{\rm Bl}\propto 1-b$. Of 
course, there is a rich area of experimental realizations with
mesoscopic superconducting devices.

We would like to mention limitations of our approach. First, 
the formula for the surface potential has been derived only for 
the magnetic field parallel to the surface. According to its 
interpretation in terms of the pairing energy we believe that it 
also applies for the perpendicular field, nevertheless, its validity 
should be tested. Second, the local approximation of the 
surface dipole requires the BCS coherence length $\xi_0$ to be 
much smaller than the GL coherence 
length $\xi$. This is satisfied at temperatures close the critical 
temperature, while one can expect sharper spatial profiles at lower 
temperatures. For this region of lower temperature, however, 
our results have to be taken only qualitatively.

\acknowledgements
We would like to thank H.~Shimada who stimulated our
interest in this problem. 
This work was supported by M\v{S}MT program Kontakt 
ME601 and GA\v{C}R 202/03/0410 and 202/04/0585, GAAV 
A1010312 grants and DAAD project D/03/44436. The 
European ESF program AQDJJ is also acknowledged.

\end{document}